\documentclass[conference]{IEEEtran}

\usepackage{xspace}
\usepackage{url}
\usepackage{graphicx}
\usepackage{cite}
\usepackage{array}
\usepackage{mdwmath}
\usepackage{mdwtab}
\usepackage{amssymb}
\usepackage[caption=false,font=footnotesize]{subfig}
\usepackage{stfloats}
\usepackage{eqparbox}
\usepackage{color}
\usepackage{amsmath}
\usepackage{amssymb}

\newcommand{\drmsim}{{DRMSim}\xspace}

\begin{document}
\title{Feasibility study on distributed simulations of BGP}

%\author{David Coudert\IEEEauthorrefmark{1},\\Aur\'{e}lien Lancin\IEEEauthorrefmark{2},\\Dimitri Papadimitriou\IEEEauthorrefmark{3},\\St\'{e}phane Perennes\IEEEauthorrefmark{5},\\Issam Tahiri\IEEEauthorrefmark{4}
%
%\bigskip
%
%\small\IEEEauthorblockA{\IEEEauthorrefmark{1}INRIA/CNRS/UNS/I3S, Route des Lucioles 2004, 06902 Sophia Antipolis, France \\Email: david.coudert@inria.fr}\\
%\IEEEauthorblockA{\IEEEauthorrefmark{2}INRIA/CNRS/UNS/I3S, Route des Lucioles 2004, 06902 Sophia Antipolis, France \\Email: aurelien.lancin@inria.fr}\\
%\IEEEauthorblockA{\IEEEauthorrefmark{3}Alcatel-Lucent Bell, Copernicuslaan 50, 2018 Antwerpen, Belgium \\Email: dimitri.papadimitriou@alcatel-lucent.be}\\
%\IEEEauthorblockA{\IEEEauthorrefmark{4}INRIA/CNRS/UNS/I3S, Route des Lucioles 2004, 06902 Sophia Antipolis, France \\Email: issam.tahiri@inria.fr}\\
%\IEEEauthorblockA{\IEEEauthorrefmark{4}INRIA/CNRS/UNS/I3S, Route des Lucioles 2004, 06902 Sophia Antipolis, France \\Email: stephane.perennes@inria.fr}}

\author{\authorblockN{David Coudert\IEEEauthorrefmark{1}, Luc Hogie\IEEEauthorrefmark{1}, Aur\'{e}lien Lancin\IEEEauthorrefmark{1}, Dimitri Papadimitriou\IEEEauthorrefmark{2}, St\'{e}phane Perennes\IEEEauthorrefmark{1}, Issam Tahiri\IEEEauthorrefmark{1} \\}
\IEEEauthorblockA{\IEEEauthorrefmark{1}INRIA, I3S(CNRS, UNS), 2004 route des Lucioles, 06902 Sophia Antipolis, France\\
\textit{Email: {(david.coudert,luc.hogie,aurelien.lancin,stephane.perennes,issam.tahiri)@inria.fr}}
\IEEEauthorblockA{\IEEEauthorrefmark{2}Alcatel-Lucent Bell, Copernicuslaan 50, 2018 Antwerpen, Belgium \\\textit{Email: dimitri.papadimitriou@alcatel-lucent.com\textsl{•}}}}}

%\author{\IEEEauthorblockN{David Coudert, Luc Hogie, Aur\'{e}lien Lancin,\\ St\'{e}phane Perennes, Issam Tahiri \\}
%\IEEEauthorblockA{INRIA, I3S(CNRS, UNS)\\
%2004 route des Lucioles, 06902 Sophia Antipolis, France\\
%\textit{Email: {david.coudert,luc.hogie,aurelien.lancin,
%stephane.perennes,issam.tahiri}@inria.fr}}
%\and
%\IEEEauthorblockN{Dimitri Papadimitriou}
%\IEEEauthorblockA{Alcatel-Lucent Bell\\
%Copernicuslaan 50, 2018 Antwerpen, Belgium\\
%\textit{Email: dimitri.papadimitriou@alcatel-lucent.com}}}

%\author{\IEEEauthorblockN{Authors Name/s per 1st Affiliation (Author)}
%\IEEEauthorblockA{line 1 (of Affiliation): dept. name of organization\\
%line 2: name of organization, acronyms acceptable\\
%line 3: City, Country\\
%line 4: Email: name@xyz.com}
%\and
%\IEEEauthorblockN{Authors Name/s per 2nd Affiliation (Author)}
%\IEEEauthorblockA{line 1 (of Affiliation): dept. name of organization\\
%line 2: name of organization, acronyms acceptable\\
%line 3: City, Country\\
%line 4: Email: name@xyz.com}
%}

\maketitle
\IEEEpeerreviewmaketitle

\begin{abstract}

%The Autonomous System (AS)-level topology of the Internet ($\sim$ 61k ASs) is growing at a rate of about 10\% per year. The Border Gateway Protocol (BGP) starts to show its limits in terms of the number of routing table entries it can dynamically process and control. DRMSim provides the means for large-scale simulations of routing models including BGP on topologies of order of 10k nodes. Therefore, DRMSim needs enhancements to support current Internet size and even more by considering its evolution (up to 100k ASs). This paper proposes a feasibility study of the extension of DRMSim so as to support the Distributed Parallel Discrete Event paradigm. We first detail the possible distribution models and their associated communication overhead. Then, we analyze their communication overhead by executing BGP on a partitioned topology according to different scenarios. Finally, we conclude on the feasibility of such a simulator by computing the expected additional time required by a distributed simulation of BGP compared to its sequential simulation.

The Autonomous System (AS) topology of the Internet (up to 61k ASs) is growing at a rate of about 10\% per year. The Border Gateway Protocol (BGP) starts to show its limits in terms of the number of routing table entries it can dynamically process and control. Due to the increasing routing information processing and storage, the same trend is observed for routing model simulators such as DRMSim specialized in large-scale simulations of routing models. Therefore, DRMSim needs enhancements to support the current size of the Internet topology and its evolution (up to 100k ASs). To this end, this paper proposes a feasibility study of the extension of DRMSim so as to support the Distributed Parallel Discrete Event paradigm. We first detail the possible distribution models and their associated communication overhead. Then, we analyze this overhead by executing BGP on a partitioned topology according to different scenarios. Finally, we conclude on the feasibility of such a simulator by computing the expected additional time required by a distributed simulation of BGP compared to its sequential simulation.
\end{abstract}


\section{Introduction}

The Internet evolution pushes its routing system to its limits in terms of i) memory cost due to the size of the routing tables (RT); ii) communication cost or complexity; iii) Border Gateway Protocol (BGP) complexity. Routing research has investigated new routing paradigms to address these issues. However, simulation of stateful distributed routing protocols operating asynchronously becomes a real issue at large-scale (order of $10k$ nodes)\cite{rob5}. No simulator provides the means to measure and to characterize the performance and behavior of such routing protocols on large networks and to compare them with BGP on the same simulation environment. % due to the processing of each individual routing state.
For this purpose, we study the extension of \drmsim\cite{drmsim} in order to support the distribution of the routing model by partitioning the topology with respect to its properties and by extending the communication model.% in order to enable the distributed execution of the routing model.

This paper is organized as follows. After describing in Section \ref{stateoftheart} the state-of-the-art in routing model simulation and \drmsim , we detail %the distributed parallel discrete-event simulation paradigm and
two distributed models together with their associated communication overhead. In Section \ref{execution}, we describe our simulation scenarios and execution environment followed by the simulation results as well as the impact of topology partition on the communication and their analysis. Finally, we conclude on the feasibility of such a simulator in Section \ref{conclusion}.
\section{State of the art}
\label{stateoftheart}

We can distinguish three classes of routing simulators: i) protocol simulators dedicated to the performance measurement and analysis of the routing protocol (procedures and format) at the microscopic level such as NS\cite{ns} and SSFNet\cite{SSFNET}; ii) configuration simulators dedicated to the simulation of BGP protocol specifics like SimBGP\cite{SimBGP} and C-BGP\cite{Quoitin05}; iii) routing model simulators like DRMSim\cite{drmsim} that do not execute the protocol low level procedures but their abstraction. Designed for the investigation of the performance of dynamic routing models on large-scale networks, these simulators allow execution of different routing models and enable comparison of their resulting performance.

%\begin{table}[t]
%\begin{center}
%\begin{tabular}{|l|ll|} 
%\hline
%Protocol simulators & NS ~\cite{ns} & SSFNet ~\cite{SSFNET} \\
%\hline
%Configuration simulators & SimBGP~\cite{SimBGP} & C-BGP ~\cite{Quoitin05} \\
%\hline
%(routing) model simulators & DRMSim~\cite{drmsim} & \\
%\hline
%\end{tabular}
%\caption{Existing simulators categories}
%\label{simulators}
%\end{center}
%\end{table}

Routing model simulators require specification of an abstract procedural model, data model, and state model sufficiently simple to be effective on large-scale networks but still representative of the actual protocol execution. However, incorporating (and maintaining up to date) routing state information is technically challenging because of the amount of memory required to store the data associated to each state. 
%In practice, processing of individual routing states impedes the execution of large-scale simulations. 
 
 \drmsim implements the Discrete-Event Simulation (DES) approach. It addresses the previous issue by means of efficient graph-based data structures. It allows us to defined simulation scenarios where topology dynamics like link/router failures are considered. Several models can be easly developed and integrated with their own metric model for performance measurement.
The provided set of models includes BGP%\cite{rfc4271}
, the Routing Information Protocol (RIP), and compact routing schemes such as NSR\cite{Nisse09} and AGMNT\cite{Abraham08}.

\section{Distribution models}

Implementing \drmsim as a distributed parallel discrete event simulator implies to distribute routers with their data structures but also the events execution among the different computational resources. The simulation feasibility depends mainly on the number of events to be transmitted between each logical processes (LP) and the available bandwidth between them.

The main problems observed with the non-distributed version of \drmsim are:
\begin{itemize}
\item Routing tables (RT): let $n$ be the number of routers in the network and $k$ the size of a routing entry. The memory needed for storing all the routing tables is $O(k.n^2)$.
\item Entries updates: event with the highest number of occurrences. We always start from empty routing tables. %Filling directly the routing tables could becomes tedious for some routing protocols that perform rule-based processing not strictly dependent on the topology. 
\end{itemize}

%As explained in Section \ref{BGP-optimization}, in the sequential version of \drmsim, the router $R_{source}$ sending an update only add the indexes of these entries from its indexed routing table; on the opposite side, the router $R_{target}$ concerned by this update reads those entries directly from the routing table of $R_{source}$. Such scheme is difficult to maintain in a distributed simulation knowing that $R_{source}$ and $R_{target}$ may not be in the same LP.

Routers and BGP peering sessions are distributed among the different logical processes. We model the system by a graph $G(V,E)$, where the set of vertices $V$ is partitioned and every partition is managed by a logical process. The edges internal to a partition are known only by the corresponding LP. For edges with end-points in distinct partitions, there are two possible solutions.

\begin{figure}[b]
\center
\includegraphics[width=0.7\columnwidth]{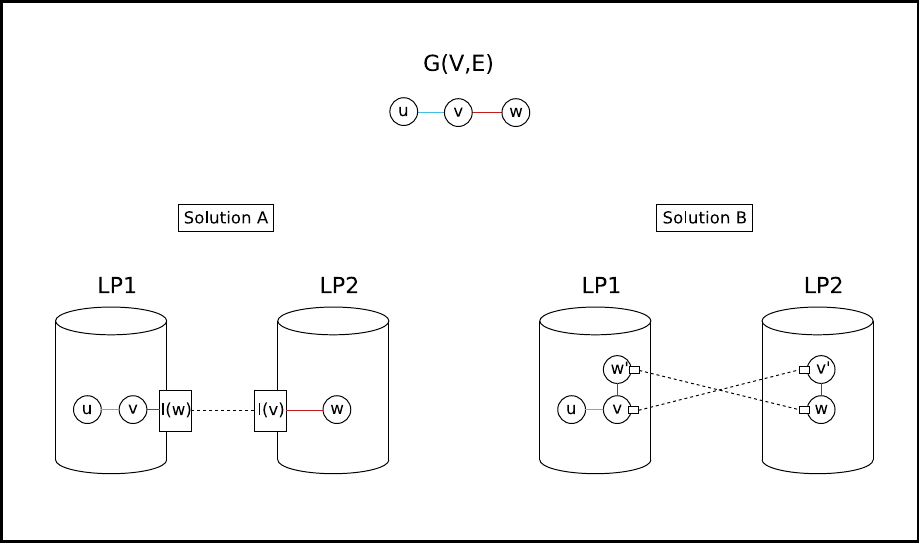}
\caption{Two solutions to allocate edges with end-points on different LPs}
\label{edges}
\end{figure}

\begin{itemize}
\item \textit{Solution A}: update events including the corresponding entries are transmitted between LPs.
\item \textit{Solution B}: boundary edge end-points are duplicated, implying that the original vertices (as an example $v$ and $w$ in Fig.\ref{edges} must synchronize their copies ($v'$ and $w'$). 
\end{itemize}

The advantage of \textit{Solution B} compared to \textit{Solution A}, is that only modified RT entries are sent, thus reducing the communication cost. However, duplication of vertices (routers) may be very harmful if the cuts resulting from the partitioning algorithm have many edges, reducing in turn the memory gain.

\subsection{Partitioning algorithms and complexity}
\label{partition}

We assume that BGP routing updates constitute the main problem to solve, other type of events lead to negligible effects. For convenience, we use the following notations:
\begin{itemize} 
\item $G(V,E) \triangleq$ the graph representing the network topology;
\item The vertices of the graph are partitioned: $V=\cup_{i=1}^K V_i$;
\item $N_{V'}(v) \triangleq$ the neighbors set of a vertex $v$ on the subgraph induced by $V'$;
\item $ME(v) \triangleq$ the set of modified RT entries at the router modeled by $v$ ;  
\item $e(v,V') = 1$ if $v$ is not in $V'$ and has a neighbor in $V'$, and 0 otherwise;
\item $Esize \triangleq$ the average size of a RT entry.
\end{itemize}

\textit{Solution A} implies no internal communication overhead, but leads to $\sum_{i=1}^k \sum_{v \in V_i} ( |N_{V \setminus V_i}(v)|.|ME(v)| )$ communication between LPs. In \textit{Solution B}, as boundary nodes are duplicated, there are $\sum_{i=1}^k \sum_{v \in V_i} [ (\sum_{j=1}^k e(v,V_j)).|ME(v)| ]$ internal updates, but the communication between LPs is negligible since we only need to send an update event containing the identifiers of both routers. There is no memory overhead in \textit{Solution A} compared to the $|V|.\sum_{i=1}^k \sum_{v \in V_i} [ (\sum_{i=1}^k e(v,V_j)) ]$ bits that are needed in \textit{Solution B} to store the duplicated boundary nodes.

%Tables \ref{tab1} and \ref{tab2} represent updates induced in terms of communication and memory overhead. Results on implementation of Solution A or Solution B are differentiated. Two types of events may occur: internal updates (blue in Figure \ref{edges}) and external updates (red in Figure \ref{edges}).

%\begin{table}[t]
%%\begin{center}
%%\begin{tabular}{|c|l|} 
%\begin{tabular}{|p{1,2cm}|p{6,8cm}|} 
%\hline
%Solution A & no memory overhead\\
%\hline
%Solution B & $Esize.|V|.\sum_{i=1}^k \sum_{v \in V_i} [ (\sum_{i=1}^k e(v,V_j)) ]$ \\
%\hline
%\end{tabular}
%\caption{Memory overhead implied by updates}
%\label{tab2}
%%\end{center}
%\end{table}

We use mixed integer programs (MIP) to compute optimal bipartitions of the topology by minimizing the communication overhead. In \textit{Solution A}, we assign to every edge $uv \in E(G)$ a weight $W_{uv}$ that approximates the number of entries exchanged on this edge. The sum of weights over all edges having end points in different subsets is minimized. In \textit{Solution B}, we assign to every vertex $v \in V(G)$ a weight $W_v$ that approximates $|ME(v)|$. In this case, the sum of weights over all vertices having at least one neighbor in the foreign subset is minimized.

\section{Execution}
\label{execution}

Our objective is to determine the communication overhead of BGP caused by a distributed parallel implementation of \drmsim. We count the number of BGP update messages exchanged at the boundary edges of the topology partitions. The communication time overhead is derived by estimating the transmission and the propagation time for a single BGP update message between two LPs.

\subsection{Simulation scenarios}

We first consider MinRouteAdvertisementInterval (MRAI) = 0s in our experiments. It represents an upper bound on the amount of communication between BGP routers. Indeed, in the Internet, the default MRAI value is set to 30s in order to limit the message rate between BGP peers.

The BGP peering session establishment delay together with the update propagation delay between routers play a major role in the amount of transmitted updates. Three scenarios have been elaborated:
\begin{itemize}
\item \textit{Scenario 1}: considers BGP peering sessions establishment before the start of updates exchanges. Once sessions are established, received updates are executed in their scheduled order. %This scenario simulates a network where all routers have already established BGP sessions with their neighbors. Communication delay of updates between routers is negligible.
\item \textit{Scenario 2}: considers \textit{Scenario 1}. However, upon reception, updates are executed in a random order simulating highly random communication delay.
\item \textit{Scenario 3}: after one peering session establishment, the resulting updates are executed. Then, the next scheduled BGP peering session is established. %This scenario simulates the arrival of routers one by one in the network, waiting for their convergence before adding a new node.
\end{itemize}

These scenarios are executed once on topologies of 2.5k, 3k, 3.5k, 4k, 4.5k and 5k nodes. These topologies are generated according to the Generalized Linear Preferential (GLP) model parameterized as specified in \cite{Bu2002}.

\subsection{Execution environment}

%Due to the huge amount of required memory to simulate BGP on topologies of order of 5k nodes with MRAI sets to 30s, 

Sequential simulations of BGP to derive inter-partitions communication have been executed with \drmsim on an Intel Xeon 3.20Ghz with 64GB of RAM. Execution of \textit{Scenario 1} took 52 minutes (min), \textit{Scenario 2} 48min and \textit{Scenario 3} 16min on topologies of 5k nodes.

\subsection{Simulation results} 

We first measure to obtain reference values, the number of BGP updates with their respective number of entries on non-partitioned topologies according to the different scenarios. 

\begin{table}[t]
\caption{Number of update entries}
\label{non-partitioned-table}
\begin{tabular}{lllllll} 
& 2.5k & 3k & 3.5k & 4k & 4.5k & 5k \\
\hline
\multicolumn{7}{l}{Scenario 1 ($\times 10^6$)}\\
\ No partition & 24.6 & 36.1 & 50.1 & 65.0 & 83.1 & 102.4 \\
\ Sol A on bipartition & 2.6 & 3.8 & 5.1 & 6.7 & 9.4 & 10.5 \\
\ Sol B on bipartition & 1.4 & 2.2 & 2.9 & 3.9 & 5 & 6.1 \\
\hline
\multicolumn{7}{l}{Scenario 2 ($\times 10^6$)}\\
\ No partition & 58.7 & 87.0 & 121.7 & 158.8 & 204.3 & 252.8 \\
\ Sol A on bipartition & 6.1 & 9.0 & 12.3 & 16 & 20.4 & 25.2 \\
\ Sol B on bipartition & 2.8 & 4.4 & 6.2 & 8 & 10.2 & 12.6 \\
\hline
\multicolumn{7}{l}{Scenario 3 ($\times 10^6$)}\\
\ No partition & 33.5 & 49.0 & 68.4 & 88.0 & 111.7 & 138.8 \\
\ Sol A on bipartition & 3.3 & 4.6 & 6.4 & 8.3 & 10.3 & 12.8 \\
\ Sol B on bipartition & 2.6 & 3.9 & 5 & 6.9 & 8.6 & 10.9 \\
\hline
\end{tabular}
\end{table}

\subsubsection{BGP reference results on non partitioned topologies} 
\label{non-partitioned}

From Table \ref{non-partitioned-table} on non partitioned topologies, the number of BGP update entries increases drastically. If the communication overhead between LPs behaves similarly, then the time to perform simulations becomes excessive. This observation shows that setting MRAI value to 0s leads to detrimental effects in terms of routing convergence time. We also observe that the increase of BGP update entries over the topology size is linear in their root square, allowing us to extrapolate this number for the topology size of interest (100k nodes).

\subsubsection{BGP results on bipartitioned topologies}
\label{partitioned}

We compute optimal bipartitions of the topologies according to the methodology presented in Section \ref{partition} with Solution A and B. As optimal bipartitions, we know that the measured communication give us a lower bound on the overhead. 

In \textit{Solution A}, for the three scenarios and a topology of 5k routers, around 10\% of the total number of updated entries transit between partitions. In \textit{Solution B}, we observe for the three scenarios that the number of transmitted entries improves compared to \textit{Solution A}. Only 5.6\% of the reference updated entries have to be transmitted with \textit{Scenario 1}, 5\% for \textit{Scenario 2} and 8\% for \textit{Scenario 3}.

\subsubsection{Communication overhead}

We measure the total size of updated entries transmitted for the different scenarios and topologies. The average size of each entry is 1 for \textit{Scenario 1} and between 2.8 and 4.5 for \textit{Scenarios 2} and \textit{Scenario 3}. We also need to determine the time to transmit an updated entry between two LPs. Considering 1Gbps link using TCP/IP, we obtain an average of 0.26ms for each packet (over 1000 packets sent one after the other) to reach its destination. 

With \textit{Solution A} applied to \textit{Scenario 1} on a topology of 5k routers, we can expect 45min to transmit all updated entries between the LPs. By extrapolation, we can expect around 91min of overhead for a topology of 10k routers and 15hours for 100k routers. For the two other scenarios executed on a topology of 5k routers, \textit{Scenario 2} cumulates an overhead of 110min and \textit{Scenario 3} remains in the same order of time as \textit{Scenario 1}. By extrapolating \textit{Scenario 2} overhead, we can expect about 4h of overhead for a topology of 10k routers and about 36hours for 100k routers.

%These results show that scenarios where perturbations are spatially distributed and the number of perturbed states limited compared to the total number of routing states can be accommodated with reasonable time overhead. On the contrary. scenarios constructed on spatial randomization of small perturbations (Scenario 2) or large number of small perturbations (Scenario 3) significantly increase this overhead. Assuming that nodes would be interconnected by 10Gbps links. one would still reach about 3 hours of cumulated time overhead for a topology of 100k routers. With Solution B applied to Scenarios~1 and~2. the number of update entries is reduced by a factor of 2 compared to Solution~A. Scenario 3 allows less improvement as only 2\% of the entries are saved. 

\section{Conclusion}
\label{conclusion}

The expansion of the Internet and its dynamics require to handle the evolutions of its routing system for the next decade. Moving \drmsim to a Distributed Parallel Discrete Event simulator seems to provide a promising technique to make abstraction of the topology size although it induces communication overhead between logical processes. Therefore, this paper quantifies the expected additional time needed to distribute the simulation of BGP on topologies representative of this growth, i.e., 10 to 100k. For this purpose, we have identified BGP updates as the main cause of communication between nodes. We computed the number of updates and their respective number of entries for different BGP execution scenarios. Then, we have computed the best bipartition of each topology - by using the previous measured number of updates - to derive the amount of communication (the overhead) between partitions. It appears that for Scenario 1, distributing the simulation between partitions requires an additional time ranging from 91min (10k nodes) to 15hours (100k nodes). As shown by the Scenario 2 and Scenario 3, this additional time clearly depends on the execution scenario and can reach up to 36hours for topologies of 100k nodes. Decreasing the number of updates can be achieved by increasing the MRAI time to a value higher than 0.  %Indeed as observed in \cite{Premore01}, there is no rate limiting when MRAI=0. 
As the MRAI increases, the number of BGP updates decreases significantly (in the order of factor 10), until the convergence time reaches an average minimum. The critical issue becomes thus to determine the optimal MRAI value, for which the decrease in update rate would not increase the convergence time while limiting the amount of additional required memory. In order to further reduce the overhead, we have proposed to synchronize boundary nodes between partitions. By applying this method, we were able to save half of the overtime when considering Scenario 1 and 2. Consequently, such a distributed parallel simulator seems thus feasible, leading in turn to the next step of our work being the extension of our routing model simulator \drmsim according to the proposed distribution models.

\end{document}